\documentclass[aps,prc,superscriptaddress,twoside,twocolumn,nofootinbib,showpacs,floatfix]{revtex4-1}
\usepackage{amsmath,amssymb}
\usepackage{graphicx}
\usepackage{subfigure}
\usepackage[colorlinks]{hyperref}
\pdfpagewidth=\paperwidth
\pdfpageheight=\paperheight
\allowdisplaybreaks
\usepackage{newtxtext, newtxmath}
\usepackage{braket}

\begin{document}

\title{\boldmath Nucleon-nucleon interaction in a chiral SU(3) quark model revisited}

\author{F. Huang}
%\email[Email: ]{huangfei@ucas.ac.cn}
\affiliation{School of Nuclear Science and Technology, University of Chinese Academy of Sciences, Beijing 100049, China}
\affiliation{Institute of High Energy Physics, Chinese Academy of Sciences, Beijing 100049, China}

\author{W. L. Wang}
\email[Email: ]{wangwenling@buaa.edu.cn}
\affiliation{School of Physics and Nuclear Energy Engineering, Beihang University, Beijing 100191, China}

\date{\today}

\begin{abstract}
A dynamical investigation of the nucleon-nucleon ($NN$) interaction by using the resonating group method (RGM) in a chiral SU(3) quark model  has been revisited. The considered quark-quark interaction includes, besides the one-gluon exchange (OGE) and the phenomenological confinement potential, the nonet scalar and pseudoscalar meson exchanges derived from the spontaneous SU(3) chiral symmetry breaking. The physical consistency requirement that the wave functions of single baryons satisfy the minimums of the Hamiltonian has been strictly imposed in determination of the model parameters. The calculated masses of the octet and decuplet baryon ground states, the binding energy of the deuteron, and the $NN$ scattering phase shifts up to a total angular momentum $J=6$ are in satisfactory agreement with the experiments.
\end{abstract}



\maketitle

\section{Introduction}  \label{Sec:Introduction}

Hadrons are composed of quarks and gluons. It is thus an exciting challenge to understand the phenomena of hadron physics directly from these fundamental degrees of freedom. Despite the progress made in understanding the consequence of quantum chromodynamics (QCD), the theory of strong interactions, the complexity of this theory in its non-perturbative region forces us to employ QCD inspired models in study of the hadron structures and the hadron-hadron interactions. Among these models, the constituent quark model has shown to be quite successful in describing the single baryon properties and the nucleon-nucleon ($NN$) and hyperon-baryon ($YN$) interactions \cite{Oka:1980,Harvey:1981,Faessler:1982,Shimizu:1984,Straub:1988,Baruer:1990,Valcarce:1994,Zhang:1994,Zhang:1997,Entem:2000,Fujiwara:2001,Mota:2002,Dai:2003,Ping:2009}. In Refs.~\cite{Huang:2004,Huang:2004-2,Huang:2005,Huang:2007,Huang:2008}, progress has also been made in understanding the kaon-nucleon ($KN$) and antikaon-nucleon (${\bar K}N$) interactions in a chiral constituent quark model. On the hadron level, the $NN$ interaction has been well described in effective field theory \cite{Epelbaum:2009,Entem:2015}.

In the constituent quark model study of $NN$ interaction, the one-gluon exchange (OGE) is found to be one of the most important sources of the short-range repulsion \cite{Oka:1980,Harvey:1981,Faessler:1982,Shimizu:1984,Straub:1988,Baruer:1990,Valcarce:1994,Zhang:1994,Zhang:1997,Entem:2000,Fujiwara:2001,Mota:2002}. Therefore, to get a proper understanding of the $NN$ short-range interaction mechanism on a quark level, one  needs a credibly determination of the coupling strengths of OGE. In earlier quark model investigations, the OGE coupling constants are usually determined by the mass differences of $N-\Delta$ and $\Lambda-\Sigma$, where the masses of single baryons are calculated as the averaged values of the Hamiltonian with the spacial wave functions of constituent quarks described by Gaussian wave functions. The assumption behind this strategy of the OGE couplings determination is that the harmonic oscillator size parameters in the Gaussian wave functions are the same for all single baryons. Under such an assumption, the matrix elements of the kinematic energy and the confinement potential for $N$ will be the same as those for $\Delta$. A similar situation applies to $\Lambda$ and $\Sigma$. Consequently, the mass differences of $N-\Delta$ and $\Lambda-\Sigma$ only come from the one-boson exchanges (OBEs) generated by the quark and chiral-field coupling and the OGE. One can then fix the OGE couplings by the mass differences of $N-\Delta$ and $\Lambda-\Sigma$ with the parameters in OBEs being predetermined. 

The problem for choosing the same size parameter in Gaussian wave functions for all single baryons is that the masses of baryon ground states are not guaranteed to be the minimums of the Hamiltonian, contradictory to the variational principle. In other words, the wave functions chosen for single baryons are not consistent with the model Hamiltonian. Physically, it is hard to understand why different baryons, e.g. $N$ and $\Delta$, or $\Sigma$ and $\Sigma^*$, have exactly the same sizes although their Hamiltonians are different due to their different quantum numbers. In earlier quark model calculations, usually the nucleon is set to be the minimum of the Hamiltonian by a particular choice of the values for other model parameters, e.g. the parameters in the confinement potential. One then needs to be very careful when extends the model from the study of $NN$ interaction to other baryon-baryon ($BB$) systems. There may be cases that one needs to introduce additional channels to lower the energy of the considered $BB$ system. These channels might not be physical ones, but are partially needed to change the internal wave functions of the single baryons. Due cautions should be taken in explaining the configuration structure of any bound $BB$ states obtained in such cases. In Refs.~\cite{Ohta:1982,Liu:1982}, Ohta {\it et al.} and Liu found that a stability condition of the nucleon should be satisfied to make a meaningful discussion of the $NN$ interaction. It is natural to expect that such an observation also holds for other $BB$ systems.

In this work, we reinvestigate the $NN$ interaction in the framework of resonating group method (RGM) within a chiral SU(3) quark model. The quark-quark interaction includes the OGE, the phenomenological confinement potential, and the nonet scalar and pseudo-scalar meson exchanges generated from the spontaneous SU(3) chiral symmetry breaking. The major difference between the present work with earlier quark model calculations is that the quark-quark interaction employed in the present work is constrained to describe the energies of single baryons, the binding energy of deuteron, and the $NN$ scattering phase shifts in a rather consistent manner without introducing any additional parameters. Specifically, in the present work, the harmonic oscillator size parameters for constituent quarks are not treated as predetermined parameters and taken to be the same for all single baryons. Instead, they are determined by variational method in calculation of the energies of single baryon ground states, which ensures that all single baryons are minimums of the Hamiltonian. The model parameters in Hamiltonian are then adjusted to simultaneously match the calculated energies of single baryons, the binding energy of the deuteron, and the predicated $NN$ scattering phase shifts with their experimental values. We mention that this is the first chiral quark model investigation where the octet and decuplet baryon ground states and the $NN$ interactions are handled in a consistent manner. It is expected that the results will be more reliable when one extends the present model from the study of $NN$ interaction to other $BB$ systems, especially to the search for dibaryons, which is expected to be done in our next step work. 

The paper is organized as follows. In the next section, we review the main aspects of the chiral SU(3) quark model, the description of the baryon ground states, the formulation of RGM, and the determination of the model parameters. The results for masses of octet and decuplet baryon ground states, the binding energy of deuteron, and the $NN$ scattering phase shifts are shown and discussed in Sec.~\ref{Sec:results}. Finally, the summary and conclusions are drawn in Sec.~\ref{Sec:summary}.

\section{Formulation}

\subsection{The chiral SU(3) quark model} \label{Sec:chQM}

The idea and details of the chiral SU(3) quark model can be found in Refs.~\cite{Zhang:1997,Huang:2004,Huang:2004-2}. Here we just present the main features of this model.

The quark and chiral field interaction Lagrangian in the flavor SU(3) case can be obtained by a linear generalization of the SU(2) linear $\sigma$-model, which gives
\begin{equation}  \label{eq:L_ch}
{\cal L}_I^{\rm ch} = - g_{\rm ch} \bar{\psi} \left( \sum^{8}_{a=0} \sigma_a \lambda^a + i \gamma_5 \sum^{8}_{a=0} \pi_a \lambda^a  \right) \psi.
\end{equation}
Here $\psi$ is the quark field, $\pi_a$ and $\sigma_a$ $(a=0,1,...,8)$ are nonet pseudoscalar and scalar fields, $\lambda^{a}$ is the Gell-Mann matrix of the flavor SU(3) group, and $g_{\rm ch}$ the quark and chiral-field coupling constant. Clearly, this Lagrangian is invariant under the infinitesimal chiral SU(3)$_L\,\times\,$SU(3)$_R$ transformation. In practice, a form factor $F(\boldsymbol{q}^{2})$ will be inserted into the vertices of quark and chiral field coupling to describe the chiral-field structure, 
\begin{equation}  \label{eq:FF}
F(\boldsymbol{q}^{2})=\left( \frac{\Lambda^2}{\Lambda^2 + \boldsymbol{q}^2}\right)^{1/2},
\end{equation}
with the cutoff mass $\Lambda$ indicating the chiral symmetry breaking scale \cite{amk91,abu91,emh91}. Equations (\ref{eq:L_ch}) and (\ref{eq:FF}) result in the following chiral field induced potentials between the $i$th and $j$th quarks:
\begin{equation}
V^{\rm ch}_{ij} = \sum_{a=0}^8 V^{\sigma_a}_{ij} + \sum_{a=0}^8 V^{\pi_a}_{ij},   \label{eq:V_ch}
\end{equation}
with
\begin{align}
V^{\sigma_a}_{ij} = &~ - C(g_{\rm ch}, m'_{\sigma_a}, \Lambda) Y_1(m'_{\sigma_a}, \Lambda, r_{ij}) \left(\lambda^a_i \lambda^a_j\right) \nonumber \\
&~ + V^{\sigma_a}_{\rm ls}(\boldsymbol{r}_{ij}),   \label{eq:V_q-q-sigma}  \\[5pt]
V^{\pi_a}_{ij} = &~ C(g_{\rm ch}, m'_{\pi_a}, \Lambda) \frac{m'^{\,2}_{\pi_a} c^a_{ij}}{48} Y_3(m'_{\pi_a},\Lambda,r_{ij})  \nonumber \\
&~ \times \left(\boldsymbol{\sigma}_i \cdot \boldsymbol{\sigma}_j\right) \left(\lambda^a_i \lambda^a_j\right)  + V^{\pi_a}_{\rm ten}(\boldsymbol{r}_{ij}),   \label{eq:V_q-q-pi}
\end{align}
and
\begin{align}
V^{\sigma_a}_{\rm ls}(\boldsymbol{r}_{ij}) = &~ - C(g_{\rm ch}, m'_{\sigma_a}, \Lambda) \frac{m'^{\,2}_{\sigma_a} s^a_{ij}}{8} Z_3(m'_{\sigma_a}, \Lambda, r_{ij})   \nonumber \\
&~ \times  \left[\boldsymbol{L} \cdot \left(\boldsymbol{\sigma}_i + \boldsymbol{\sigma}_j\right)\right]   \left(\lambda^a_i \lambda^a_j\right),  \\[5pt]
V^{\pi_a}_{\rm ten}(\boldsymbol{r}_{ij}) = & ~ C(g_{\rm ch}, m'_{\pi_a}, \Lambda) \frac{m'^{\,2}_{\pi_a} c^a_{ij}}{48} H_3(m'_{\pi_a}, \Lambda, r_{ij})  \nonumber\\
&~ \times \left[3\left(\boldsymbol{\sigma}_i \cdot \hat{\boldsymbol{r}}_{ij}\right)\left(\boldsymbol{\sigma}_j \cdot \hat{\boldsymbol{r}}_{ij}\right)- \boldsymbol{\sigma}_i \cdot \boldsymbol{\sigma}_j\right] \left(\lambda^a_i \lambda^a_j\right),
\end{align}
where
\begin{align}
C(g_{\rm ch},m,\Lambda) = &~ \frac{g^2_{\rm ch}}{4\pi} \frac{\Lambda^2}{\Lambda^2-m^2} m,  \\[5pt]
Y_1(m,\Lambda,r) = &~ Y(mr) - \frac{\Lambda}{m} Y(\Lambda r),  \label{x1mlr}  \\[5pt]
Y_3(m,\Lambda,r) = &~ Y(mr) - \left(\frac{\Lambda}{m}\right)^3 Y(\Lambda r),  \\[5pt]
Z_3(m,\Lambda,r) = &~ Z(mr) - \left(\frac{\Lambda}{m}\right)^3 Z(\Lambda r),  \\[5pt]
H_3(m,\Lambda,r) = &~ H(mr) - \left(\frac{\Lambda}{m}\right)^3 H(\Lambda r),  \\[5pt]
Y(x ) = &~ \frac{1}{x}e^{-x},  \\[5pt]
Z(x) = &~ \left(\frac{1}{x}+\frac{1}{x^2}\right)Y(x),  \\[5pt]
H(x) = &~ \left(1+\frac{3}{x}+\frac{3}{x^2}\right)Y(x),  \\[5pt]
c^a_{ij} = &~ \left\{\begin{array}{lcl} \dfrac{4}{m_i m_j},  && (a=0,1,2,3,8) \\[5pt]  \dfrac{\left(m_i + m_j\right)^2}{m_i^2 m_j^2},  && (a=4,5,6,7) \end{array} \right.   \\[5pt]
s^a_{ij} = &~ \left\{\begin{array}{lcl} \dfrac{1}{m_i^2}+\dfrac{1}{m_j^2},  && (a=0,1,2,3,8) \\[12pt]   \dfrac{2}{m_i m_j},   && (a=4,5,6,7) \end{array} \right.
\end{align}
and $m_i$ and $m_j$ are masses of the $i$th and $j$th constituent quarks. $m'_{\sigma_a}$ and $m'_{\pi_a}$ are related to the mass of scalar meson, $m_{\sigma_a}$, and the mass of pseudoscalar meson, $m_{\pi_a}$, by
\begin{align}  \label{eq:mass_S}
m_{\sigma_a}' &= \left\{\begin{array}{lcl} m_{\sigma_a},   && (a=0,1,2,3,8) \\[6pt]    \sqrt{m_{\sigma_a}^2-\left(m_i-m_j\right)^2},   && (a=4,5,6,7) \end{array} \right.  \\[5pt]
m_{\pi_a}' &= \left\{\begin{array}{lcl} m_{\pi_a},   && (a=0,1,2,3,8) \\[6pt]    \sqrt{m_{\pi_a}^2-\left(m_i-m_j\right)^2}.   && (a=4,5,6,7) \end{array} \right.  \label{eq:mass_PS}
\end{align}
The relations for $a=4,5,6,7$ in Eqs.~(\ref{eq:mass_S}) and (\ref{eq:mass_PS}) come from an explicit treatment of the mass difference of $u(d)$ and $s$ quarks in the derivation of potentials of Eqs.~(\ref{eq:V_q-q-sigma}) and (\ref{eq:V_q-q-pi}) for $\kappa$- and $K$-exchange from the quark and chiral field interaction Lagrangian of Eq.~(\ref{eq:L_ch}).

For pseudoscalar meson exchanges, the fact that the physical $\eta$ and $\eta$' are mixing states of $\eta_0$ and $\eta_8$ is considered in the calculation:
\begin{equation}
\left\{ \begin{array}{l} \eta = \eta_8 \cos\theta^{PS} - \eta_0 \sin\theta^{PS}, \\[5pt]  \eta' = \eta_8 \sin\theta^{PS} + \eta_0 \cos\theta^{PS}, \end{array} \right.
\end{equation}
with the mixing angle $\theta^{PS}$ taken to be the empirical value $\theta^{PS} = -23^\circ$.

Apart from the potentials from OBEs given in Eq.~(\ref{eq:V_ch}), to study the hadron structure and hadron-hadron dynamics, one still needs the potential from OGE,
\begin{align}
V^{\rm OGE}_{ij}  = &~ \frac{g_i g_j}{4} \left(\boldsymbol{\lambda}^c_i \cdot \boldsymbol{\lambda}^c_j\right) \left\{\frac{1}{r_{ij}}-\frac{\pi}{2} \delta(\boldsymbol{r}_{ij}) \left[ \frac{1}{m^2_i}+\frac{1}{m^2_j} \right.\right. \nonumber \\
&~ \left. + \left. \frac{4}{3}\frac{1}{m_i m_j} \left( \boldsymbol{\sigma}_i \cdot \boldsymbol{\sigma}_j \right) \right] \right\}  + V^{\rm OGE}_{\rm ls}({\boldsymbol r}_{ij})  + V^{\rm OGE}_{\rm ten}({\boldsymbol r}_{ij}),
\end{align}
with
\begin{align}
V^{\rm OGE}_{\rm ls}({\boldsymbol r}_{ij}) = &~ -\frac{g_ig_j}{4} \left(\boldsymbol{\lambda}^c_i \cdot \boldsymbol{\lambda}^c_j\right) \frac{m_i^2+m_j^2+4m_im_j}{8m_i^2 m_j^2}\frac{1}{r^3_{ij}}  \nonumber \\
&~ \times  \left[ {\boldsymbol{L} \cdot \left( \boldsymbol{\sigma}_i + \boldsymbol{\sigma}_j \right)} \right],  \\[5pt]
V^{\rm OGE}_{\rm ten}({\boldsymbol r}_{ij}) = &~ -\frac{g_ig_j}{4} \left(\boldsymbol{\lambda}^c_i \cdot \boldsymbol{\lambda}^c_j\right) \frac{1}{4m_i m_j}\frac{1}{r^3_{ij}}  \nonumber \\
&~ \times \left[3\left(\boldsymbol{\sigma}_i \cdot \hat{\boldsymbol{r}}_{ij}\right)\left(\boldsymbol{\sigma}_j \cdot \hat{\boldsymbol{r}}_{ij}\right)- \boldsymbol{\sigma}_i \cdot \boldsymbol{\sigma}_j\right],
\end{align}
and a phenomenological confinement potential of which the frequently used linear and quadratic types are considered in the present work,
\begin{equation}
V_{ij}^{\rm conf} = \left\{\begin{array}{lcl}  -\left(\boldsymbol{\lambda}_{i}^{c}\cdot\boldsymbol{\lambda}_{j}^{c}\right) \left(a_{ij}r_{ij} +a_{ij}^{0}\right),  && {\rm (Model~I)}  \\[8pt] -\left(\boldsymbol{\lambda}_{i}^{c}\cdot\boldsymbol{\lambda}_{j}^{c}\right) \left(a_{ij}r_{ij}^2 +a_{ij}^{0}\right).  && {\rm (Model~II)}
%  \\[6pt]  -\left(\boldsymbol{\lambda}_{i}^{c}\cdot\boldsymbol{\lambda}_{j}^{c}\right) \left(a_{ij} {\rm Erf}[c_0 r_{ij}] +a_{ij}^{0}\right), 
\end{array}\right.   \label{eq:conf}
\end{equation}

The total Hamiltonian for a multiquark system can then be written as
\begin{align}
H =&~ \sum_{i=1}^N \left(m_i + \frac{\boldsymbol{p}_i^2}{2m_i} \right) - \frac{\left(\sum_{i=1}^N \boldsymbol{p}_i\right)^2}{2\sum_{i=1}^N m_i} + \sum_{j > i = 1}^N \left(V^{\rm conf}_{ij} \right. \nonumber \\
&~ + \left. V_{ij}^{\rm OGE} + V_{ij}^{\rm ch}  \right), \label{eq:hamiltonian}
\end{align}
with $\boldsymbol{p}_i$ being the three-momentum of the $i$th quark, and $N$ being the number of quarks for the system considered.

\subsection{Baryon ground states} \label{Sec:baryon}

The antisymmetrized wave function for decuplet or octet ground state baryon $B$ in spin-flavor-color-orbit space can be written as
\begin{equation}  \label{wf-3q}
\Psi_B = \left\{ \begin{array}{lcl} \Phi_B^{\rm S} \chi^{\rm S}_B \zeta_B,   &&    \mbox{(decuplet)}   \\[5pt]  \displaystyle\frac{1}{\sqrt 2}\left(\sum_{{\rm MX} = {\rm MS}, {\rm MA}} \Phi_B^{\rm MX}  \chi_B^{\rm MX} \right)  \zeta_B,    &&   \mbox{(octet)} \\
\end{array}\right.
\end{equation}
where $\Phi_B$, $\chi_B$ and $\zeta_B$ are wave functions of baryon $B$ in orbit-flavor, spin and color spaces, respectively, with the superscripts S, MX denoting symmetric and mixed-symmetric under interchange of any pair of quarks in the corresponding space. Note that the baryon wave functions in orbit and flavor spaces are coupled together, since the mass of strange quark $m_s$ is different from that of up (down) quark $m_u$ ($m_d$),
\begin{align}  \label{wf-3q-fo1}
\Phi_{B}^{\rm S} &= \sum_{f_1f_2f_3} {C_B^{\rm S}}(f_1f_2f_3) \, \psi(\boldsymbol{r}_1\boldsymbol{r}_2\boldsymbol{r}_3; f_1f_2f_3)  \Ket{ f_1f_2f_3 }, \\[5pt]
 \Phi_{B}^{\rm MX} &= \sum_{f_1f_2f_3} {C_B^{\rm MX}}(f_1f_2f_3) \, \psi(\boldsymbol{r}_1\boldsymbol{r}_2\boldsymbol{r}_3; f_1f_2f_3) \Ket{f_1f_2f_3},   \label{wf-3q-fo2}
\end{align}
where $f_i$ is the flavor of the $i$th quark, $C_B(f_1f_2f_3)$ the SU(3) Clebsch-Gordon coefficient in flavor space, and $\psi(\boldsymbol{r}_1\boldsymbol{r}_2\boldsymbol{r}_3; f_1f_2f_3)$ the orbit wave function for three quarks with flavor contents $f_1f_2f_3$. The orbit wave function is determined by the variational method with the trial wave function taken to be a product of Gaussian functions,
\begin{equation}
\psi(\boldsymbol{r}_1\boldsymbol{r}_2\boldsymbol{r}_3; f_1f_2f_3) = \prod_{i=1}^3 \, \left(\frac{1}{\pi b^2_i}\right)^{3/4}\mbox{exp}\left[-\frac{1}{2b_i^2}\boldsymbol{r}_i^2\right].    \label{eq:Gaussian}
\end{equation}
With the assumption that the harmonic oscillator frequency is the same for $u$, $d$ and $s$ quarks which ensures that the three-quark center-of-mass motion is irrelevant to the calculation, the size parameter for strange quark, $b_s$, is related to that for up quark, $b_u$, by
\begin{equation}
b_s = \sqrt{\frac{m_u}{m_s}} b_u.
\end{equation}
In earlier quark model calculations, $b_u$ is treated as a predetermined parameter and taken as the same for all single baryons. In the present work, $b_u$ for each baryon will be obtained by minimize the corresponding baryon mass from the model calculation, i.e.
\begin{equation}
\frac{\partial}{\partial{b_u}} \Braket{\Psi_B | H | \Psi_B } = 0,  \label{eq:baryon-mass-variation}
\end{equation}
which ensures that each single baryon is the solution of the Hamiltonian of Eq.~(\ref{eq:hamiltonian}).

\subsection{RGM for $NN$ system} \label{Sec:RGM}

The following Jacobi coordinates are defined to construct the total wave function of the $NN$ system:
\begin{align}
\boldsymbol{\xi}_1 &= \boldsymbol{r}_2 - \boldsymbol{r}_1,  \\[3pt]
\boldsymbol{\xi}_2 &= \boldsymbol{r}_3 - \frac{\boldsymbol{r}_1 + \boldsymbol{r}_2}{2},  \\[3pt]
\boldsymbol{\xi}_3 &= \boldsymbol{r}_5 - \boldsymbol{r}_4,  \\[3pt]
\boldsymbol{\xi}_4 &= \boldsymbol{r}_6 - \frac{\boldsymbol{r}_4 + \boldsymbol{r}_5}{2},  \\[3pt]
\boldsymbol{r} &= \frac{\boldsymbol{r}_1 + \boldsymbol{r}_2 + \boldsymbol{r}_3}{3} - \frac{
\boldsymbol{r}_4 + \boldsymbol{r}_5 + \boldsymbol{r}_6}{3},
\end{align}
with $\boldsymbol{\xi}_i$ ($i=1-4$) being the internal coordinates for two nucleon clusters, and $\boldsymbol{r}$ bing the relative coordinate of two clusters.

Following the cluster model calculations \cite{mka77,mok81}, the RGM wave function is written as
\begin{equation}
\Psi_{NN} = {\cal A} \left[ \Psi_N(\boldsymbol{\xi}_1,\boldsymbol{\xi}_2) \Psi_N(\boldsymbol{\xi}_3,\boldsymbol{\xi}_4) \chi_{\rm rel}(\boldsymbol{r})\right]_{ST},
\end{equation}
with $\chi_{\rm rel}(\boldsymbol{r})$ being the trial wave function of relative motion of two nucleon clusters, and $ST$ being the total spin and isospin of the $NN$ system. $\Psi_N$ is the internal wave function of each single cluster, taken to be the same as that from Eq.~(\ref{wf-3q}). The wave function of total center-of-mass motion is stripped off as it is irrelevant to the calculation. The symbol $\cal A$ is the antisymmetrizing operator for interchange any pair of constituent quarks between two clusters,
\begin{equation}
{\cal A} = 1 - \sum_{i=1}^3 \sum_{j=4}^6 P_{ij},
\end{equation}
where $P_{ij}$ is the permutation operator of the $i$th and $j$th quarks that are from different clusters. Substituting $\Psi_{NN}$ into the following projection equation
\begin{equation}
\Braket{ \delta\Psi_{NN} | H - \left(2E_N + E_{\rm rel}\right) | \Psi_{NN} } =0,  \label{eq:projection}
\end{equation}
with $E_N$ and $E_{\rm rel}$ being the inner energy of nucleon and the relative energy between two nucleon clusters, respectively, one gets the RGM equation for the unknown relative motion wave function
\begin{equation}
\int {\cal L}(\boldsymbol{r}, \boldsymbol{r}') \chi_{\rm rel}(\boldsymbol{r}') d\boldsymbol{r}' =0,
\end{equation}
with
\begin{equation}
{\cal L}(\boldsymbol{r}, \boldsymbol{r}') = {\cal H}(\boldsymbol{r}, \boldsymbol{r}') - \left(2E_N + E_{\rm rel}\right) {\cal N}(\boldsymbol{r}, \boldsymbol{r}'),
\end{equation}
where the Hamiltonian kernel $\cal H$ and normalization kernel
$\cal N$ can, respectively, be calculated by
\begin{align}
\left\{
       \begin{array}{c}
          {\cal H}(\boldsymbol{r}, \boldsymbol{r}') \\
          {\cal N}(\boldsymbol{r}, \boldsymbol{r}')
       \end{array}
\right\}
= \bigg \langle \left[\Psi_N(\boldsymbol{\xi}_1,\boldsymbol{\xi}_2)
\Psi_N(\boldsymbol{\xi}_3, \boldsymbol{\xi}_4) \delta(\boldsymbol{r}- \boldsymbol{r}_{NN}) \right]_{ST}  \bigg | \nonumber \\
\times 
\left\{
      \begin{array}{c}
          H \\
          1
       \end{array}
\right\}
\bigg | {\cal A} \left[\Psi_N(\boldsymbol{\xi}_1,\boldsymbol{\xi}_2)
\Psi_N(\boldsymbol{\xi}_3, \boldsymbol{\xi}_4) \delta(\boldsymbol{r}'-\boldsymbol{r}_{NN})\right]_{ST} \bigg\rangle.
\end{align}

In practical calculation, the unknown wave function for two-cluster relative motion $\chi_{\rm rel}(\boldsymbol{r})$ is projected into partial waves,
\begin{equation}
\chi_{\rm rel}(\boldsymbol{r}) = \sum_L \frac{1}{r} \chi_{\rm rel}^L(r) Y_{LM}(\hat{\boldsymbol{r}}).
\end{equation}
For a bound state problem, the $L$-wave relative wave function $\chi_{\rm rel}^L(r)$ is expanded as
\begin{equation}
\chi_{\rm rel}^L(r) = \sum_{i=1}^n c_i u^L(r, S_i),
\end{equation}
with
\begin{align}
u^L(r, S_i)  \equiv &~ 4\pi r \left(\frac{3}{2\pi b^2_u}\right)^{3/4} \mbox{exp}\left[-\frac{3}{4b_u^2}\left(r^2 + S_i^2\right)\right]  \nonumber \\  &~ \times i_L\left(\frac{3}{2b_u^2}rS_i\right),
\end{align}
where $S_i$ is called the generate coordinate, and $i_L$ the $L$th modified spherical Bessel function. The information about the unknown wave function of two-cluster relative motion is now exhibited by the coefficients $c_i$'s with properly chosen values of the generate coordinates. Performing a variational procedure, one deduces the $L$th partial-wave equation for a bound-state problem,
\begin{equation} \label{eq:bound}
\sum_{j=1}^n {\cal L}^L_{ij} c_j = 0, \qquad (i=1,...,n)
\end{equation}
with
\begin{align}
& {\cal L}^L_{ij} = \int u^L(r, S_i){\cal L}^L(r, r') u^L(r',
S_j)rr'drdr',  \\[5pt]
& {\cal L}^L(r, r') = \int Y^*_{LM}(\hat{\boldsymbol{r}})  {\cal L}(\boldsymbol{r}, \boldsymbol{r}') Y_{LM}(\hat{\boldsymbol{r}}')d\hat{\boldsymbol{r}}d\hat{\boldsymbol{r}}'.    \label{rgmkernel}
\end{align}
Solving Eq.~$(\ref{eq:bound})$, one gets the binding energy and the corresponding wave function of the two-cluster system.

For a scattering problem, the radial part of the relative motion wave function is expanded as
\begin{equation}
\chi_{\rm rel}^L(r) = \sum_{i=1}^n c_i {\tilde u}^L(r, S_i),
\end{equation}
with
\begin{equation}
\tilde{u}^L(r, S_i) \equiv \left\{
\begin{array}{lcl}
          p_i u^L(r, S_i),   &&  (r \leq R_0) \\[8pt]
          \left[h_L^-(kr) - s_i h_L^+(kr)\right] r,   &&  (r \geq R_0)
\end{array} \right.
\end{equation}
where $h_L^{\pm}$ is the $L$th spherical Hankel functions, $k$ is the momentum of the two-cluster relative motion, and $R_0$ is a cutoff radius beyond which all the strong interactions can be disregarded. The complex parameters $p_i$ and $s_i$ are determined by the smoothness condition at $r=R_0$ and $c_i$'s satisfy $\sum_{i=1}^nc_i=1$. Performing a variational procedure, the $L$th partial-wave equation for the scattering problem can be deduced as
\begin{equation}    \label{eq:scatter}
\sum_{j=1}^{n-1} {\tilde {\cal L}}^L_{ij} c_j = {\tilde {\cal
M}}_i^L,  \qquad (i=1,...,n)
\end{equation}
with
\begin{align}
{\tilde {\cal L}}^L_{ij} &= {\tilde {\cal K}}^L_{ij}-{\tilde {\cal K}}^L_{in}-{\tilde {\cal K}}^L_{nj}+{\tilde {\cal K}}^L_{nn},  \\[5pt]
{\tilde {\cal M}}^L_i &= {\tilde {\cal K}}^L_{nn}-{\tilde {\cal K}}^L_{in},
\end{align}
and
\begin{equation}
{\tilde {\cal K}}^L_{ij}=\int {\tilde u}^L(r, S_i){\cal L}^L(r, r') {\tilde u}^L(r', S_j)rr'drdr',
\end{equation}
where the RGM kernel ${\cal L}^L(r, r')$ is defined in Eq.~$(\ref{rgmkernel})$. Solving Eq.~$(\ref{eq:scatter})$, the $S$-matrix element
$S^L$ and the phase shifts $\delta_L$ can be obtained by
\begin{equation}    \label{eq:phase}
S^L\equiv e^{2i\delta_L}=\sum_{i=1}^n c_i s_i.
\end{equation}

\subsection{Model parameters} \label{Sec:para}

\begin{table}[tbp]
\caption{\label{Table:para} Model parameters. The meson masses and the cutoff masses: $m_{\sigma'} = m_{\kappa} = m_{\epsilon}=980$ MeV, $m_{\pi}=138$ MeV, $m_K=495$ MeV, $m_{\eta}=549$ MeV, $m_{\eta'}=957$ MeV, $\Lambda=1100$ MeV. The masses of $u(d)$ quark and $s$ quark are taken to be $313$ MeV and $470$ MeV, respectively. The strengths of confinement $a_{uu}$, $a_{us}$, $a_{ss}$ are in MeV/fm for linear confinement and in MeV/fm$^2$ for quadrature confinement, respectively. The mass of $\sigma$ meson, $m_\sigma$,  and the constants $a^{0}_{uu}$, $a^{0}_{us}$ and $a^{0}_{ss}$ are in MeV. }
\begin{tabular*}{\columnwidth}{@{\extracolsep\fill}lrr}
\hline\hline
  & Model I & Model II   \\ 
  & ($r$ conf.) & ($r^2$ conf.)   \\
\hline
 $m_\sigma$  & $608$ & $625$   \\
 $g_u$     & $1.06$ & $0.98$   \\
 $g_s$     & $1.16$ & $1.07$   \\
 $a_{uu}$  & $86.5$ & $56.2$    \\
 $a_{us}$  & $90.4$ & $69.3$    \\
 $a_{ss}$  & $100.4$ & $101.3$   \\
 $a^{0}_{uu}$   & $-51.4$ & $-36.6$   \\
 $a^{0}_{us}$   & $-33.4$ & $-25.1$   \\
 $a^{0}_{ss}$  & $-12.8$ & $-14.9$   \\
\hline\hline
\end{tabular*}
\end{table}

The predetermined model parameters are: (1) the masses of mesons, which are taken to be their experimental values except for $\sigma$ meson, whose mass is treated as a parameter to be fitted by the binding energy of deuteron and the $NN$ scattering phase shifts; (2) the cutoff mass $\Lambda$ in the form factor of Eq.~(\ref{eq:FF}), which is taken to be 1100 MeV as usual \cite{Dai:2003,Huang:2004,Huang:2004-2,Huang:2005,Huang:2007,Huang:2008}; (3) the constituent $u$, $d$, $s$ quark masses, which are chosen to be $m_u=m_d=313$ MeV, $m_s=470$ MeV \cite{Zhang:1997,Dai:2003,Huang:2004,Huang:2004-2,Huang:2005,Huang:2007,Huang:2008}; (4) the quark and chiral field coupling constant, which is fixed by the relation
\begin{equation}
\frac{g_{\rm ch}^2}{4\pi} = \frac{9}{25} \frac{m_u^2}{M_N^2} \frac{g_{NN\pi}^2}{4\pi},   \label{eq:gch}
\end{equation}
with $g_{NN\pi}$ taken to be the empirical value ${g_{NN\pi}^2}/{4\pi}=13.67$. 

The left parameters are the coupling constants of OGE, the parameters in confinement potential, and the mass of $\sigma$ meson. They are adjusted to match the baryon masses calculated at the minimums of the Hamiltonian, the binding energy of deuteron, and the $NN$ scattering phase shifts with their experimental values. The fitted values of those parameters are listed in Table~\ref{Table:para}, where model I and model II refer to models with linear and quadratic confinement as presented in Eq.~(\ref{eq:conf}), respectively.

One sees from Table~\ref{Table:para} that the OGE coupling constants fixed in the present work are a little bit larger than those in earlier quark model calculations, but they are still qualitatively consistent with those usually expected in QCD at the squared momentum transfer $Q^2\sim 1$ GeV$^2$, namely $\alpha_s(Q^2=1)\sim 1$.

Note that in earlier quark model calculations, the harmonic oscillator size parameter $b_u$ in the Gaussian wave functions is also treated as a predetermined parameter, and the masses of all the octet and decuplet baryons are then calculated by using the same $b_u$ as that for nucleon. As mentioned in Sec.~\ref{Sec:Introduction}, the problem in such calculations is that the calculated energies of singe baryon ground states other than nucleon are not minimums of the Hamiltonian, contradictory to the variational principle. Therefore, cautions should be taken when one uses the same set of parameters to study the $BB$ systems other than $NN$. This issue will be further discussed in the next section.

\section{Results and discussions} \label{Sec:results}

As shown in Table~\ref{Table:para}, we have nine adjustable model parameters, namely the sigma meson mass $m_\sigma$, the OGE coupling constants $g_u$ and $g_s$, the confinement strengths $a_{uu}$, $a_{us}$, $a_{ss}$ and zero point energies $a^0_{uu}$, $a^0_{us}$, $a^0_{ss}$. With the parameter values listed in Table~\ref{Table:para}, we get the masses of baryon ground states by variational method of Eq.~(\ref{eq:baryon-mass-variation}), the deuteron binding energy by solving Eq.~(\ref{eq:bound}), the $NN$ scattering phase shifts by solving Eq.~(\ref{eq:scatter}) and calculating Eq.~(\ref{eq:phase}).

\begin{table*}[tbp]
\caption{\label{Table:B-mass} Resulted mass and size parameter of octet and decuplet baryon ground states.}
\begin{tabular*}{\textwidth}{@{\extracolsep\fill}lcccccccc}
\hline\hline
 & $N$ & $\Lambda$ & $\Sigma$ & $\Xi$ & $\Delta$ & $\Sigma^*$ & $\Xi^*$ & $\Omega$  \\ \hline
Expt. [MeV] & $939$ & $1116$ & $1193$ & $1318$ & $1232$ & $1385$ & $1533$ & $1672$ \\
Theo. [MeV] & $939$ & $1116$ & $1193$ & $1318$ & $1232$ & $1385$ & $1533$ & $1672$ \\
$b_u$ [fm] ($r$ conf.) & $0.474$ & $0.478$ & $0.507$ & $0.487$ & $0.642$ & $0.632$ & $0.615$ & $0.593$ \\
~~~~~~~~~~~~\;($r^2$ conf.) & $0.472$ & $0.473$ & $0.495$ & $0.476$ & $0.588$ & $0.578$ & $0.561$ & $0.540$ \\
\hline\hline
\end{tabular*}
\end{table*}

Table~\ref{Table:B-mass} shows our results for energies and corresponding size parameters of octet and decuplet baryon ground states. One sees that our theoretical masses calculated for all baryon ground states are consistent with their experimental values in both linear confinement and quadratic confinement models. One also observes that the size parameters in the Gaussian wave functions of Eq.~(\ref{eq:Gaussian}) are quite different for various baryons, and they are also different in models with linear or quadratic confinement.

\begin{figure*}[tbp]
{\vglue 0.1cm}
\includegraphics[width=0.9\columnwidth]{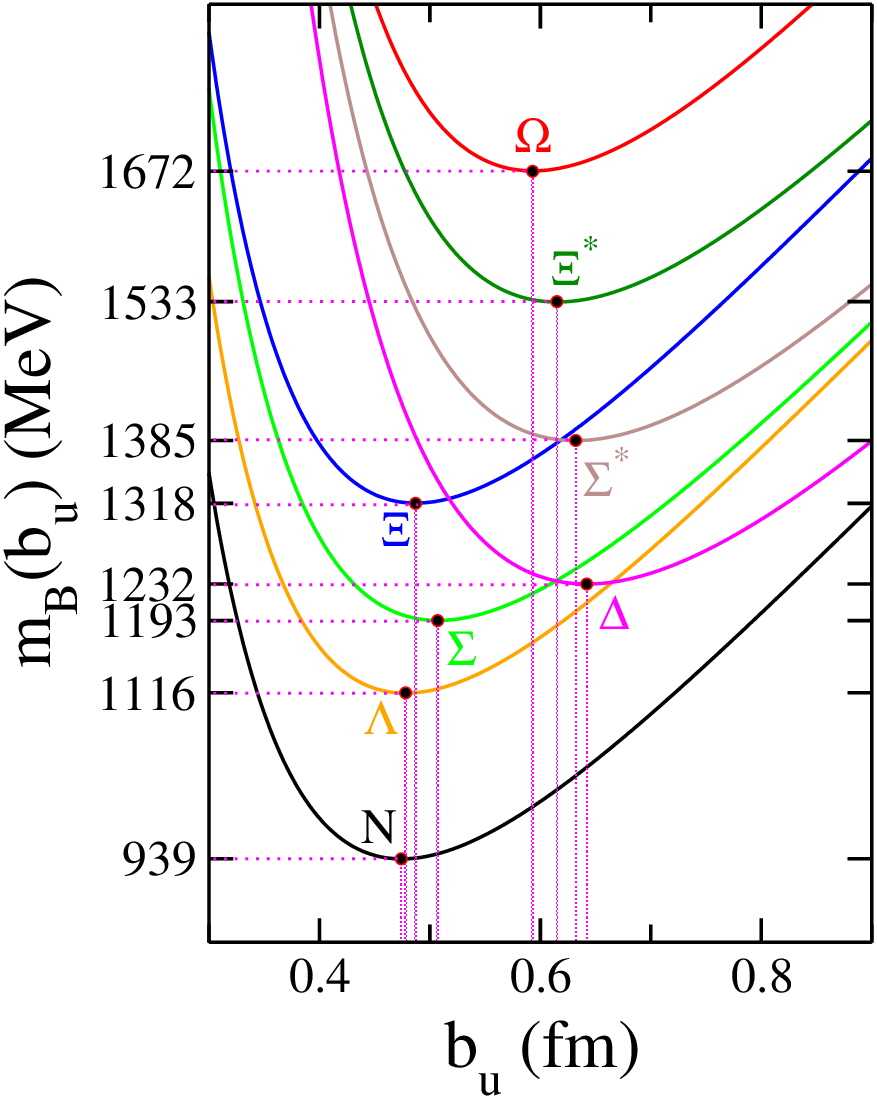} {\hglue 1.0cm}
\includegraphics[width=0.9\columnwidth]{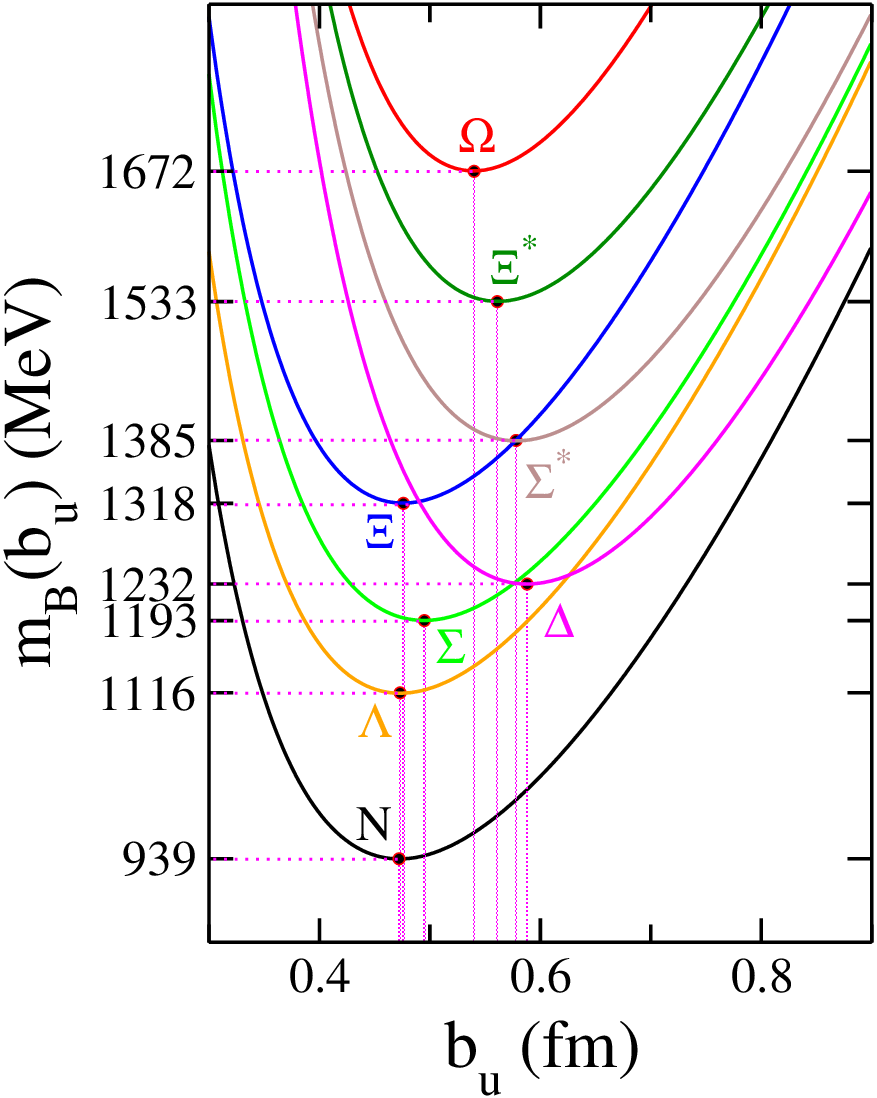}
\caption{Baryon mass $m_B$ as a function of its variational size parameter $b_u$. Left: model I (linear confinement). Right: model II (quadratic confinement). }
\label{Fig:baryon_mass_size}
\end{figure*}

\begin{table}[tbp]
\caption{\label{Table:Bind-d} Binding energy of deuteron (in MeV).}
\begin{tabular*}{\columnwidth}{@{\extracolsep\fill}ccc}
\hline\hline
 Model I & Model II  & Expt.  \\ 
  ($r$ conf.) & ($r^2$ conf.)  &  \\
\hline
$-2.215$ & $-2.218$   & $-2.224$ \\
\hline\hline
\end{tabular*}
\end{table}

\begin{figure*}[tbp]
\includegraphics[width=0.75\textwidth]{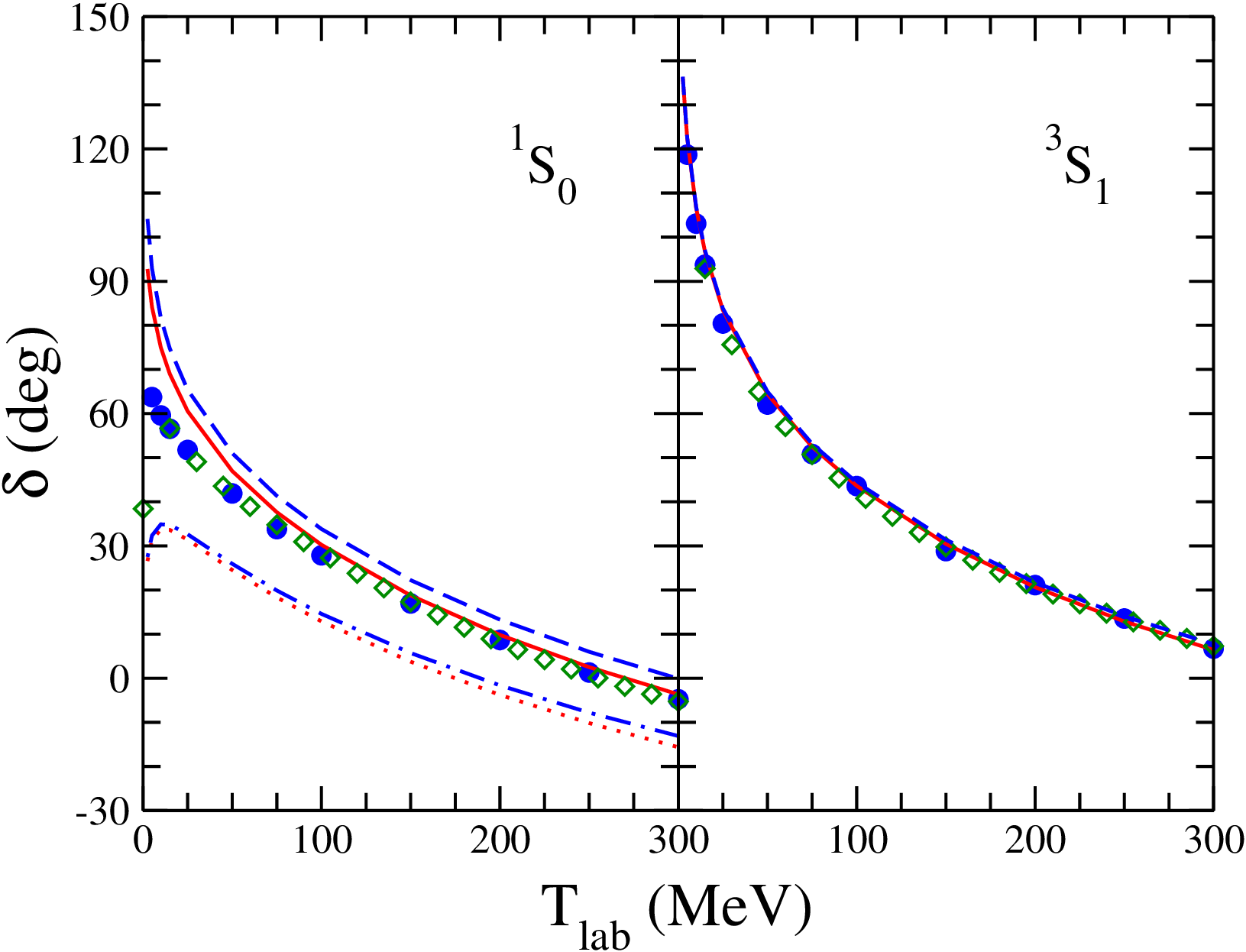}
\caption{$NN$ $S$-wave phase shifts. The red solid and blue dashed lines represent the results calculated in models with linear and quadratic confinement, respectively. For $^1S_0$ partial wave, the red dotted and blue dash-dotted lines correspond to the results calculated without considering the coupling of $NN$ $^1S_0$ to $\Delta\Delta$ $^5D_0$ in models with linear and quadratic confinement, respectively. The scattered symbols represent the results from partial wave analysis of SAID, with full circles denoting the single energy analysis and empty diamonds the energy dependent solution \cite{Workman:2016}. }
\label{Fig:S-wave}
\end{figure*}

\begin{figure*}[tbp]
\includegraphics[width=0.75\textwidth]{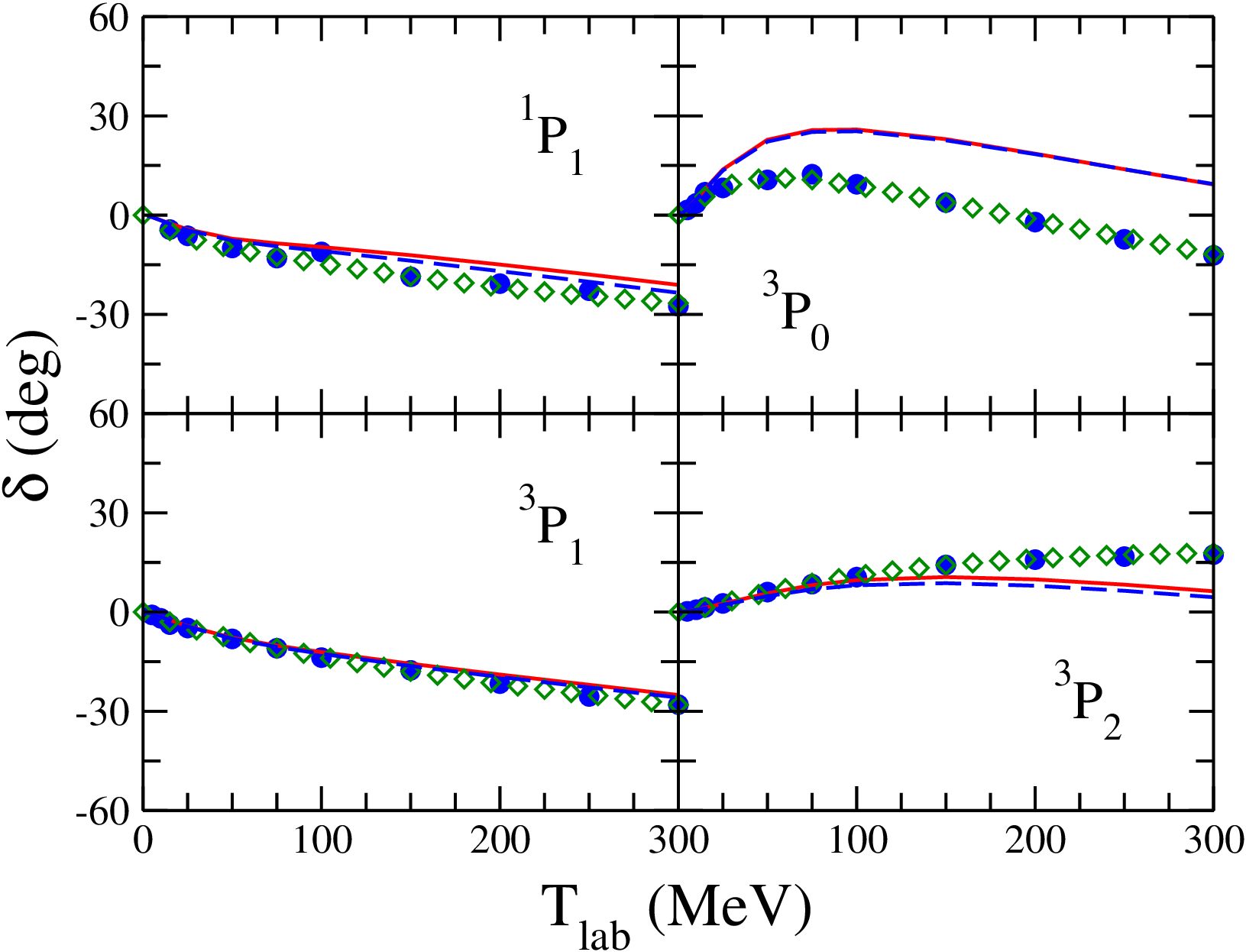}
\caption{$NN$ $P$-wave phase shifts. The notations are the same as in Fig.~\ref{Fig:S-wave}.}
\label{Fig:P-wave}
\end{figure*}

\begin{figure*}[tbp]
\includegraphics[width=0.75\textwidth]{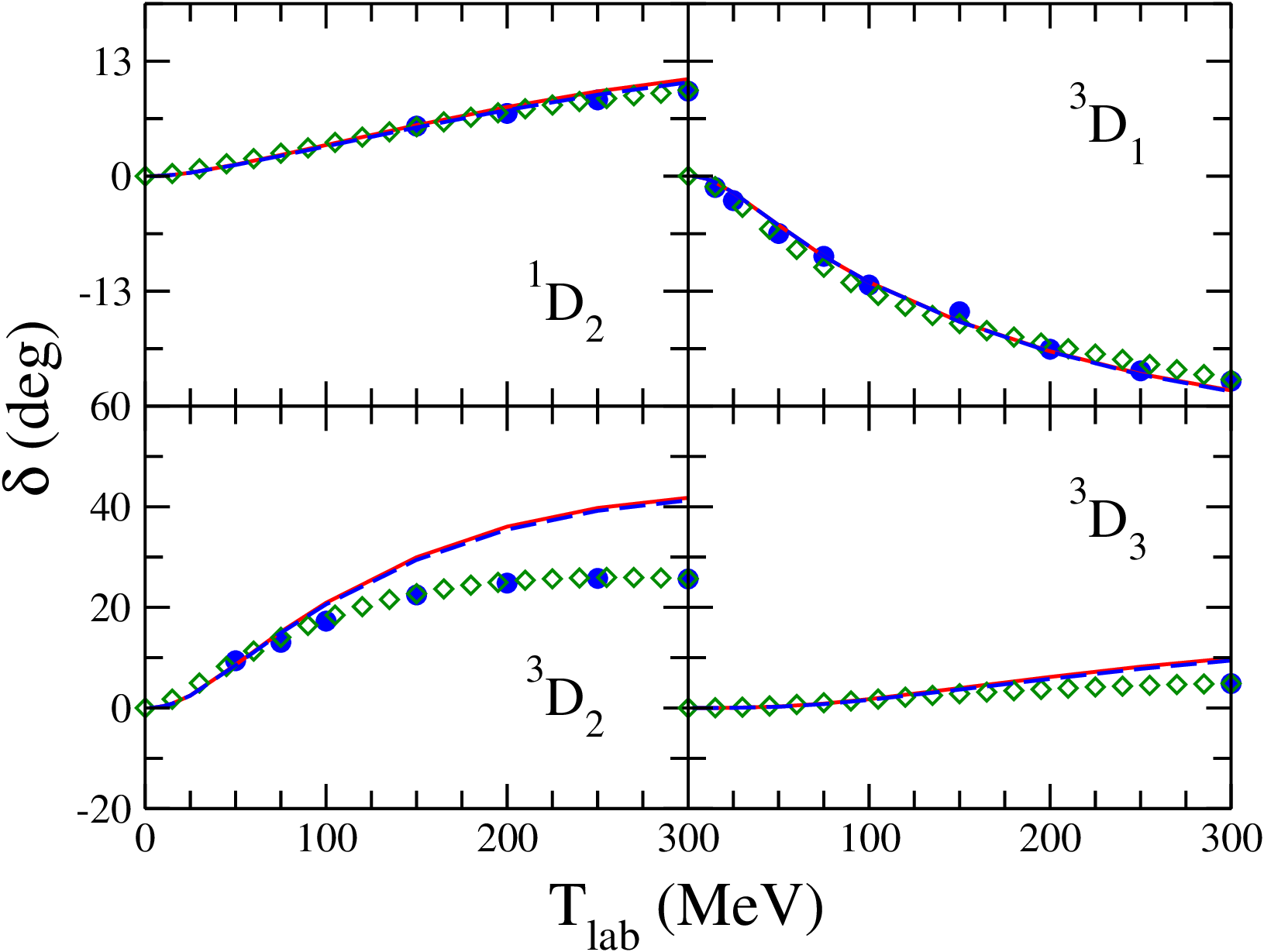}
\caption{$NN$ $D$-wave phase shifts. The notations are the same as in Fig.~\ref{Fig:S-wave}.}
\label{Fig:D-wave}
\end{figure*}

\begin{figure*}[tbp]
\includegraphics[width=0.75\textwidth]{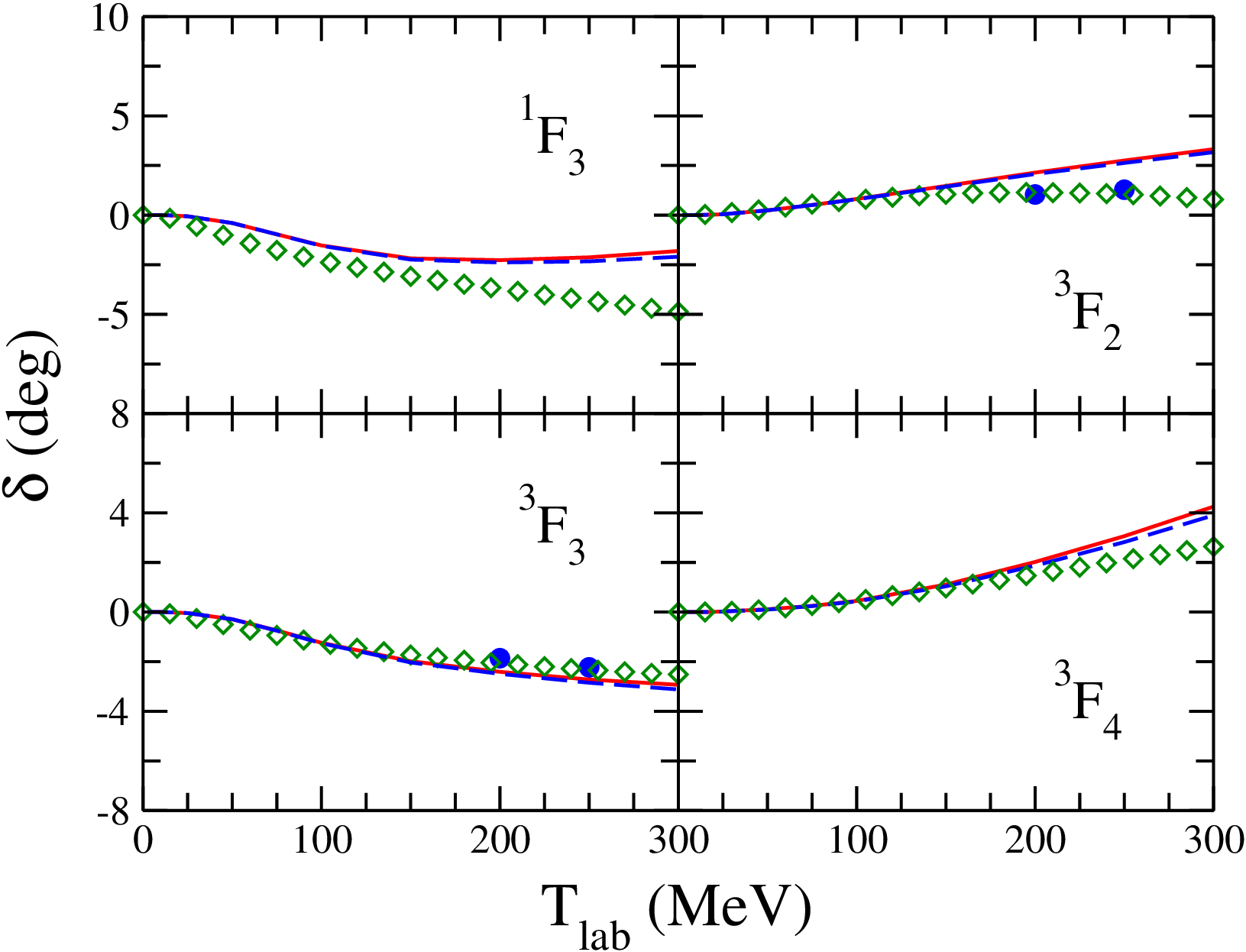}
\caption{$NN$ $F$-wave phase shifts. The notations are the same as in Fig.~\ref{Fig:S-wave}.}
\label{Fig:F-wave}
\end{figure*}

\begin{figure*}[tbp]
\includegraphics[width=0.75\textwidth]{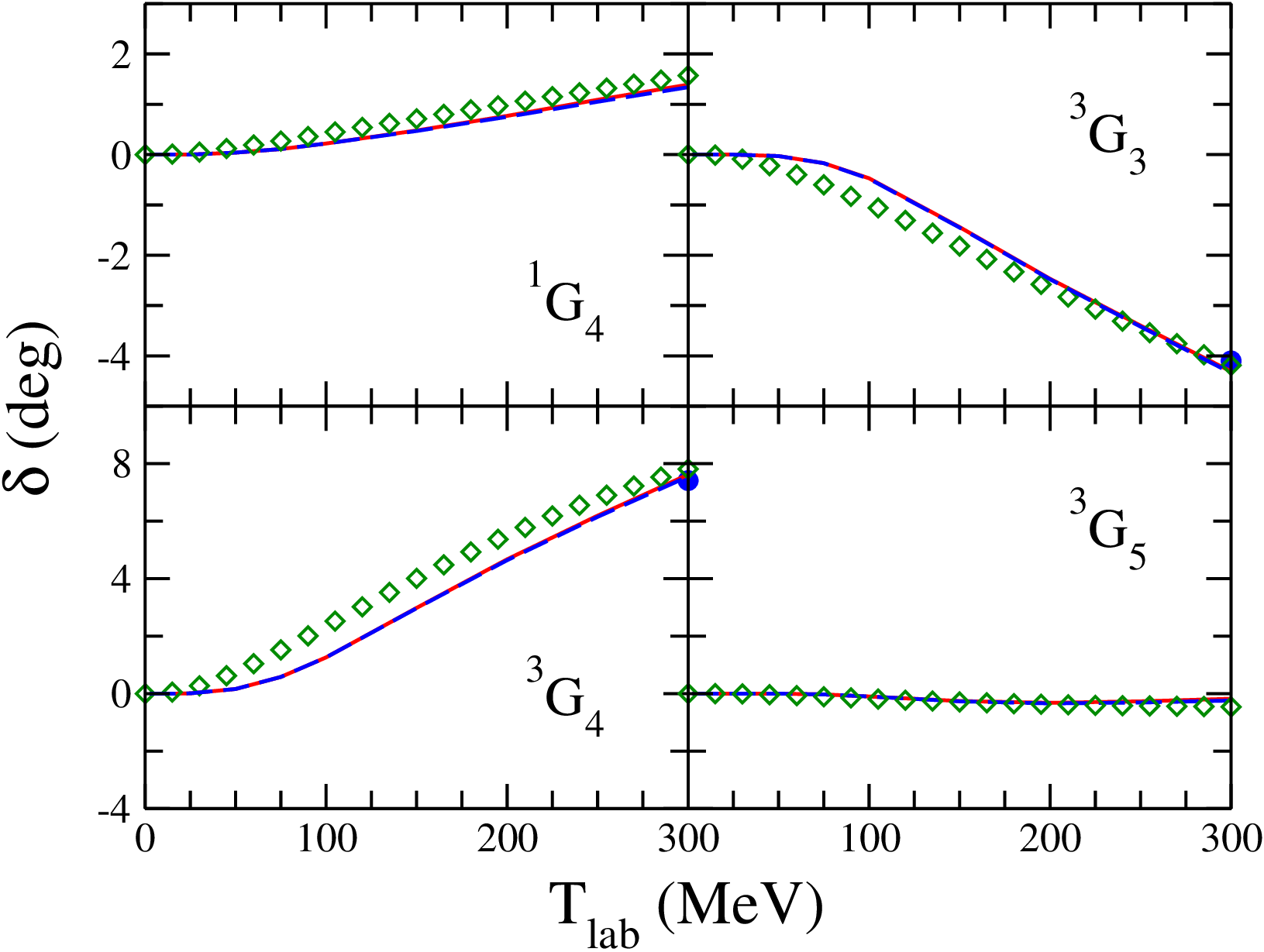}
\caption{$NN$ $G$-wave phase shifts. The notations are the same as in Fig.~\ref{Fig:S-wave}.}
\label{Fig:G-wave}
\end{figure*}

\begin{figure*}[tbp]
\includegraphics[width=0.75\textwidth]{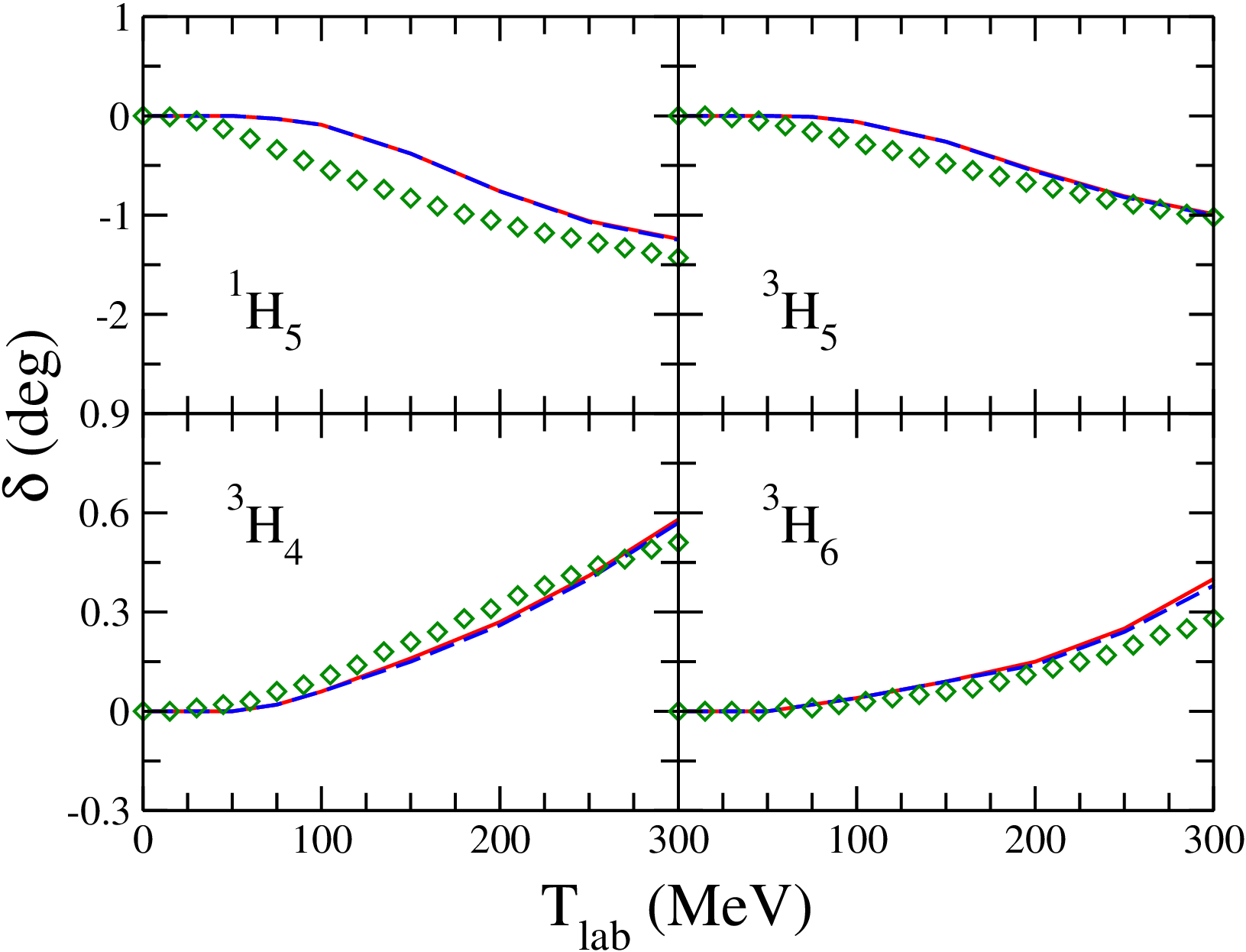}
\caption{$NN$ $H$-wave phase shifts. The notations are the same as in Fig.~\ref{Fig:S-wave}.}
\label{Fig:H-wave}
\end{figure*}

\begin{figure*}[tbp]
\includegraphics[width=0.75\textwidth]{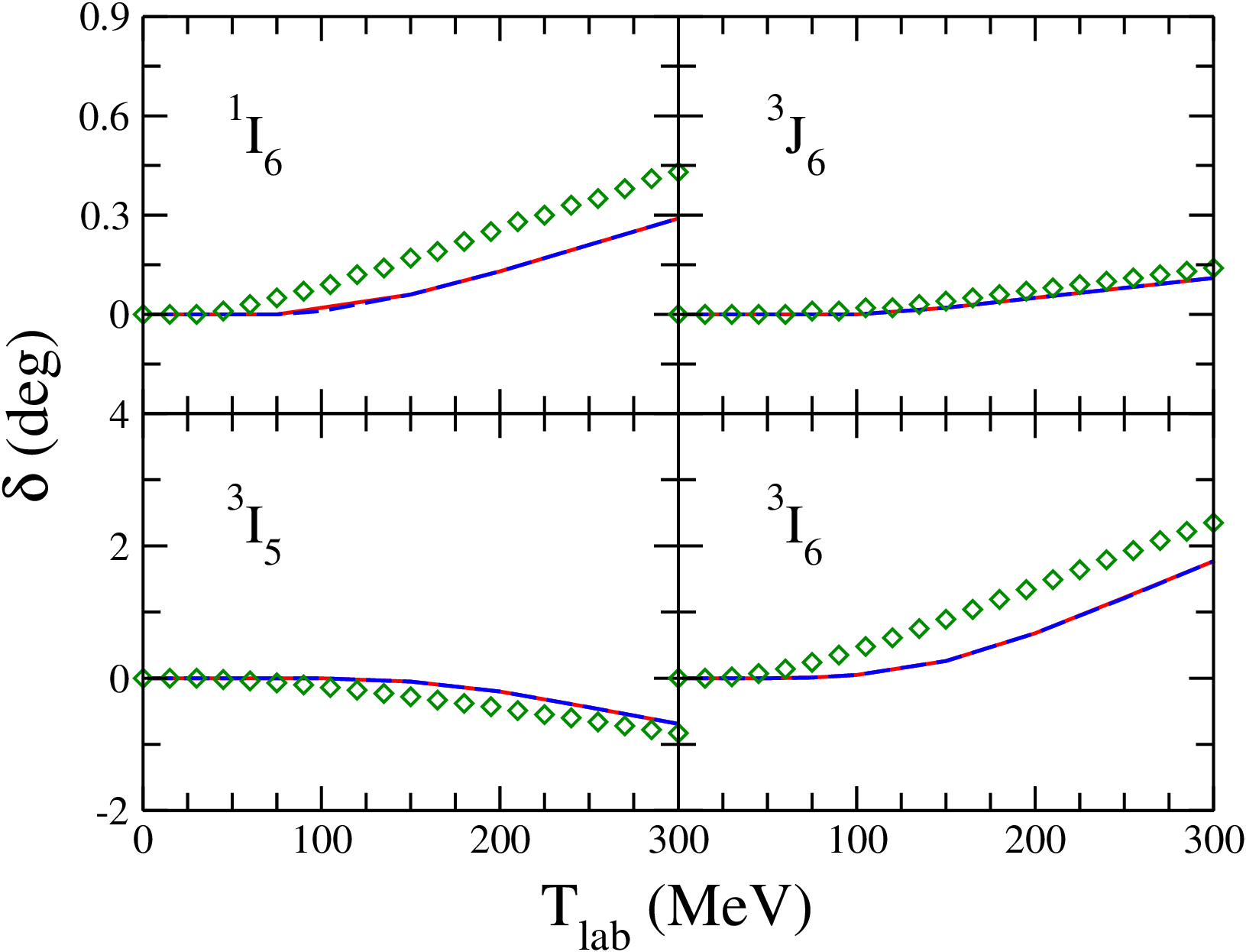}
\caption{$NN$ $I$-wave and  $^3J_6$-wave phase shifts. The notations are the same as in Fig.~\ref{Fig:S-wave}.}
\label{Fig:IJ-wave}
\end{figure*}

\begin{figure*}[tbp]
\includegraphics[width=0.75\textwidth]{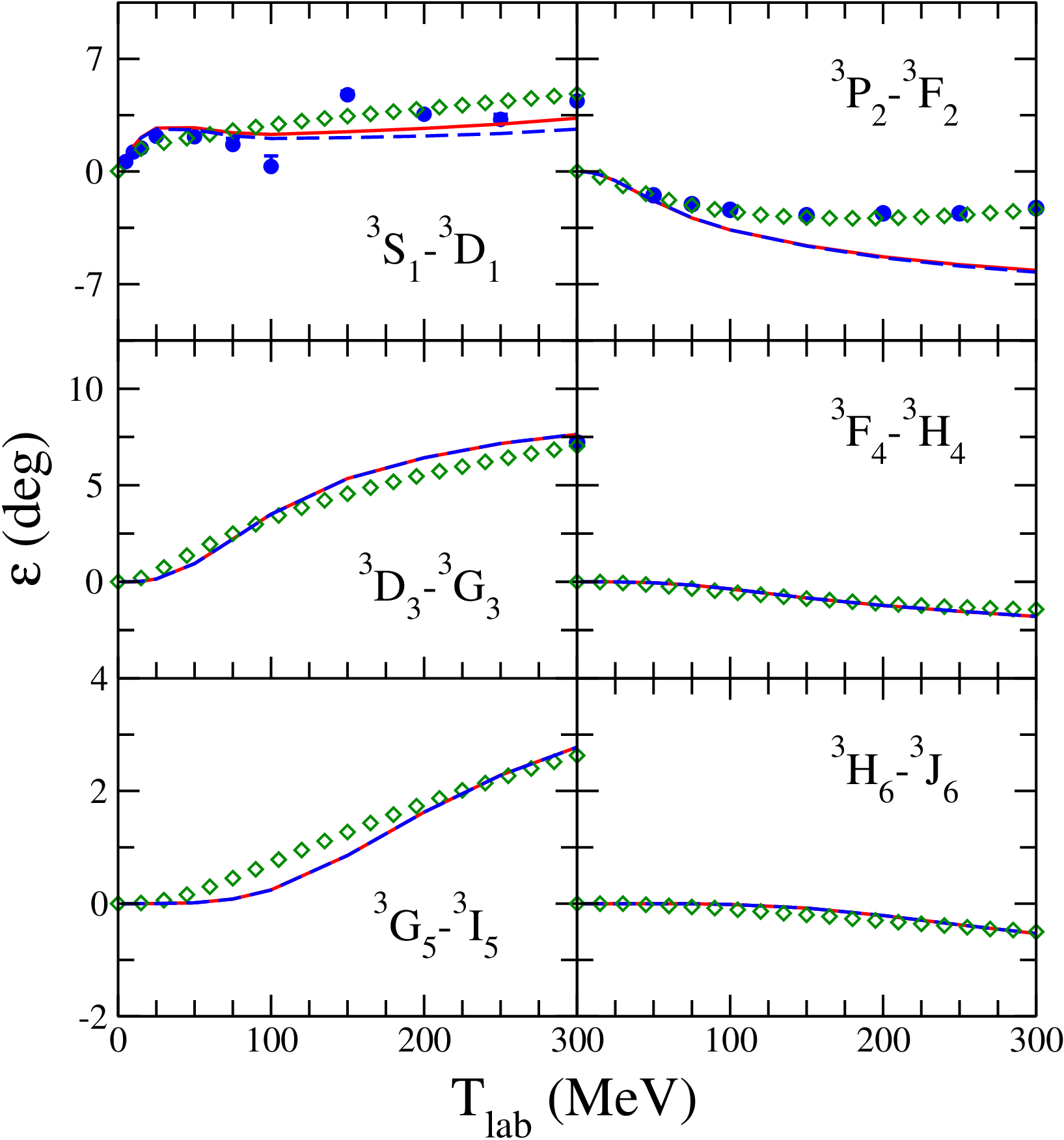}
\caption{Mixing parameters for $NN$ coupled partial waves. The notations are the same as in Fig.~\ref{Fig:S-wave}.}
\label{Fig:mixing-para}
\end{figure*}

\begin{figure*}[tbp]
\includegraphics[width=0.75\textwidth]{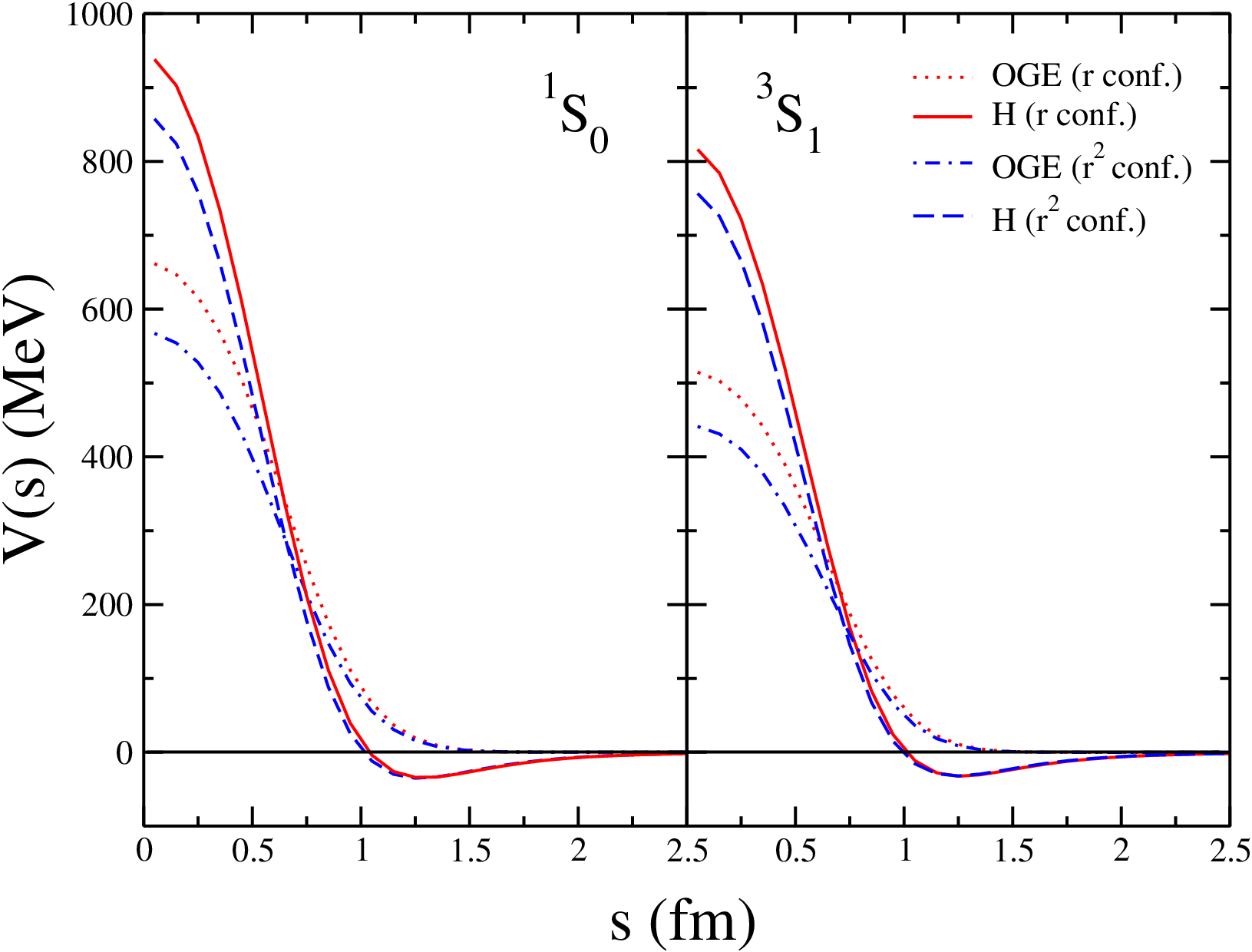}
\caption{The GCM diagonal matrix elements of OGE and Hamiltonian for $NN$ $^1S_0$ and $^3S_1$ partial waves. The red solid and blue dashed lines represent the matrix elements for Hamiltonian in models with linear and quadratic confinement, respectively. The red dotted and blue dash-dotted lines correspond to the results for OGE in models with linear and quadratic confinement, respectively.}
\label{Fig:OGE-H-S-wave}
\end{figure*}

In earlier quark model calculations, the size parameter $b_u$ is treated as a predetermined parameter and it is the same for all $N$, $\Delta$ and other single baryons. In our opinion, once the Hamiltonian is given, the baryon wave functions should be obtained in principle by solving a Schr\"odinger equation. Of course this is impractical due to the complexity of the quark-quark interacting potential [cf. Eq.~(\ref{eq:hamiltonian})]. An approximated solution could then be obtained by using Gaussians as trial wave functions and determining the baryon masses and sizes by variational method. This has, for the first time, been strictly imposed in the present work without introducing any additional parameters compared with earlier quark model calculations \cite{Zhang:1997,Dai:2003}. 

In Fig.~\ref{Fig:baryon_mass_size} we show our calculated baryon mass as a function of the variational size parameter $b_u$ for each baryon ground state. One sees that the minimums of the masses of various baryons are located at different values of the size parameter $b_u$. For octet baryons, the size parameters resulted from the linear confinement model are close to those from the quadratic confinement model. While for decuplet baryons, the size parameters resulted from the linear confinement model are much bigger than those from the quadratic confinement model. This indicates that the decuplet baryons are much more sensitive than the octet baryons to the choice of the type of the phenomenological confinement potential. One also observes that in both models, the size parameters for decuplet baryons are always much bigger than those for octet baryons, which is distinct from the earlier quark model calculations in literature, where the size parameters for all octet and decuplet baryons are taken to be the same. Under such an assumption, the OGE coupling constants are claimed to be determined by the mass differences of $N$-$\Delta$ and $\Lambda$-$\Sigma$, with the masses of all $N$, $\Delta$, $\Lambda$ and $\Sigma$ baryons calculated at the same size parameter $b_u$. In these calculations, one may adjust the parameters in the confinement potential to make the masses of $N$ and $\Lambda$ stable against the parameter $b_u$, but then the $\Delta$, $\Sigma$ and other baryons are obviously not the minimums of the Hamiltonian, as indicated by Table~\ref{Table:B-mass} and Fig.~\ref{Fig:baryon_mass_size}. When one extends the model from the study of $NN$ interaction to other $BB$ systems especially when the decuplet baryons are involved, there might be cases that one needs to introduce additional channels to lower the energy of the considered $BB$ system by changing the internal wave functions of the single baryons. Due caution should be exercised in explaining the configuration structure of any bound states obtained in such kind of calculations.

Table~\ref{Table:Bind-d} shows the binding energy of deuteron calculated in the present work with a comparison with the experimental value. One sees that our calculated values in both linear confinement model and quadratic confinement model are in good agreement with the data. This can be easily understood, since in the parameters fitting procedure, we found that this quantity is rather sensitive to the mass of $\sigma$ meson, and one thus can fine tune $m_\sigma$ to make the theoretical binding energy of deuteron close to the experimental data. 

Figures~\ref{Fig:S-wave}-\ref{Fig:IJ-wave} show the $NN$ scattering phase shifts and Fig.~\ref{Fig:mixing-para} shows the mixing parameters for $NN$ coupled partial waves up to a total angular momentum $J=6$ calculated in the present work, where the red solid and blue dashed lines represent the results calculated in models with linear confinement and quadratic confinement, respectively. The scattered symbols are results form SAID's partial wave analysis \cite{Workman:2016}, with the full circles representing their single energy analysis and the empty diamonds representing their energy dependent solution (SM16). One sees that the overall agreement of our results with SAID's partial wave analysis is satisfactory. At least they are not worse than the earlier quark model calculations, while in the present work we have put much more strict constraints on the model parameters.

Figure~\ref{Fig:S-wave} shows the $S$-wave $NN$ phase shifts. For $^3S_1$ partial wave, the theoretical phase shifts are in perfect agreement with the data. This channel corresponds to the deuteron quantum numbers, which means that the coupling to the $^3D_1$-partial wave caused by the tensor force resulted from OGE and one-pion exchange is important. Given the size parameter $b_u$ for nucleon determined by variational method of Eq.~(\ref{eq:baryon-mass-variation}), and the mass of sigma meson $m_\sigma$ fine tuned to get the deuteron binding energy, the phase shifts for this partial wave are found to be automatically in agreement with the data in the present work. The differences of the phase shifts from the linear confinement model and quadratic confinement model are found to be negligible. For $^1S_0$ partial wave, the calculated phase shifts represented by the red dotted and blue dash-dotted lines of the left subfigure indicate that there is a lack of attraction in this partial wave. It is known that the required attraction in this partial wave is supplied by the coupling to the $N\Delta$ $^5D_0$ partial wave \cite{Zhang:1997}, caused by the tensor force offered by OGE and one-pion exchange. In the present work, since $N$ and $\Delta$ have different size parameters, it is rather complicated to perform a rigorous calculation of the coupling between $NN$ $^1S_0$ and $N\Delta$ $^5D_0$ partial waves in the framework of RGM. Here for the sake of simplicity, we estimate the effect from this coupling by calculating the off-diagonal transition matrix elements at nucleon's size parameter. The resulted phase shifts are shown as red solid line (linear confinement model) and blue dashed line (quadratic confinement model) in left subfigure of Fig.~\ref{Fig:S-wave}. It is clearly seen that the coupling to the $N\Delta$ $^5D_0$ partial wave improves the $NN$ $^1S_0$ phase shifts significantly. Nevertheless, there are still noticeable discrepancies in the energy region of $T_{\rm lab} < 100$ MeV, indicating a redundant attraction obtained from the coupling to the $N\Delta$ $^5D_0$ partial wave. A more complete analysis of this partial wave requires a strict calculation of the coupling of $NN$ $^1S_0$ and $N\Delta$ $^5D_0$ partial waves in RGM with the difference of the size parameters $b_u$ for $N$ and $\Delta$ been properly taken into account. Such an investigation requires a new development of the RGM and is postponed until our future work. 

The $P$-wave $NN$ phase shifts are shown in Fig.~\ref{Fig:P-wave}. One sees that the phase shifts for $^1P_1$-wave and $^3P_1$-wave are described satisfactorily. The $^3P_0$ phase shifts are too attractive, while the $^3P_2$ phase shifts are a little bit less attractive in the energy region of $T_{\rm lab} > 150$ MeV, indicating the need for a moderate spin-orbit force, a topic deserving special treatment outside this general work.

Figure~\ref{Fig:D-wave} shows the $D$-wave $NN$ phase shifts. Apart from the $^3D_2$ partial wave, which seems too attractive in the energy region of $T_{\rm lab} > 100$ MeV, the phase shifts for other partial waves are described quite well.

The phase shifts for the $NN$ $F$-wave are shown in Fig.~\ref{Fig:F-wave}. The $^3F_3$ partial wave is well described. The $^3F_2$ and $^3F_4$ partial waves are perfectly described in the $T_{\rm lab} < 200$ MeV energy region, and they are a little bit too attractive when $T_{\rm lab} > 200$ MeV. The $^1F_3$ partial wave is a little bit too attractive in the $T_{\rm lab} > 150$ MeV energy region.

Figures~\ref{Fig:G-wave}-\ref{Fig:IJ-wave} show the results for $G$, $H$, $I$ and $J$ partial waves up to a total angular momentum $J=6$. The phase shifts for all the partial waves are described quite well, except that there are minor deviations in a few partial waves. 

Figure~\ref{Fig:mixing-para} shows the mixing parameters for the relevant $NN$ coupled partial waves. The coupling of these spin-triplet partial waves are due to the tensor forces stemming from the OGE and pseudoscalar meson exchanges. One sees that the mixing parameters for all the coupled partial waves considered in the present work are in good agreement with the experimental data, except that for the coupling of $^3P_2$-$^3F_2$, which agrees with data only at very low energies. Note that the phase shifts for the $^3P_2$ and $^3F_2$ partial waves are not quite well described in the present work, as shown in Figs.~\ref{Fig:P-wave} and \ref{Fig:F-wave}.

In a word, the $NN$ scattering phase shifts for $S$, $P$, $D$, $F$, $G$, $H$, $I$ and $J$ partial waves up to a total angular momentum $J=6$ and mixing parameters for the relevant coupled partial waves have been satisfactorily described in the present work. Note that the model has only 9 adjustable parameters as tabulated in Table~\ref{Table:para}. It is known that the confinement potential itself results in negligible contributions between two color-singlet clusters \cite{Huang:2004,Huang:2004-2,Huang:2005,Huang:2007}. Therefore, among these 9 adjustable parameters, only $g_u$, $g_s$ and $m_\sigma$ are relevant to the $NN$ scattering observables. Since $g_u$ and $g_s$ have already been determined in fitting the masses of octet and decuplet single baryons, apart from the fact that $m_\sigma$ is fine tuned to reproduce the binding energy of the deuteron, the calculation of the $NN$ scattering phase shifts for 26 partial waves and mixing parameters for the relevant coupled partial waves in the present work is almost parameter-free. With this in mind, it is fair to say that the description of the $NN$ scattering data in the present work is promising.

In earlier quark model calculations, the couplings of $g_u$, $g_s$ and the mass of sigma meson $m_\sigma$ are independent of the choice of confinement type (linear or quadratic) which is assumed to be proportional to the color-color operator as shown in Eq.~(\ref{eq:conf}). There, the masses of all single baryons are calculated at the same size parameter $b_u$. Of course by doing so the single baryons other than $N$ and $\Lambda$ are not minimums of the model Hamiltonian. In spite of that, the $N$ and $\Delta$ get the same contributions from the kinematic energy and the confinement potential, so do $\Lambda$ and $\Sigma$. Then the parameters $g_u$ and $g_s$ can be uniquely determined by the mass differences of $N-\Delta$ and $\Lambda-\Sigma$ once the quark-chiral field coupling is fixed by Eq.~(\ref{eq:gch}) and the cutoff mass is selected to be around the chiral symmetry breaking scale. The parameters in the confinement potential are just used to make the $N$ and $\Lambda$ stable against the size parameter $b_u$ and to adjust the matrix elements of Hamiltonian calculated at $b_u$ close to the empirical masses of all single baryon ground states. Since the $g_u$ and $g_s$ are independent of the choice of confinement type, and the confinement potential itself does not considerably contribute to the $NN$ interaction, the sigma meson mass $m_\sigma$ determined by fitting the $NN$ scattering data will then also be independent of the choice of confinement type. When one extends the model to study other $BB$ systems composed of two color-singlet clusters, no sizable differences will be observed when different types of confinement potential are chosen.

The situation is quite different in the present work. One sees that the parameters $g_u$, $g_s$ and $m_\sigma$ are different in linear confinement model and quadratic confinement model. The reason for that is the following. We want to describe the baryon ground states, the binding energy of deuteron, and the $NN$ scattering processes in a consistent manner. This requires the size parameter in the wave functions of each single baryon to be determined by minimization of the Hamiltonian. As can be seen in Table~\ref{Table:B-mass} and Fig.~\ref{Fig:baryon_mass_size}, the resulted size parameters are different for different baryons in both the linear confinement model and the quadratic confinement model. Therefor, unlike in the earlier quark model calculations, the kinematic energies and the confinement potential now also contribute to the mass differences of $N-\Delta$ and $\Lambda-\Sigma$ due to the unequal size parameters of these baryons. The mass differences remained for contributions of OGE are thus usually different for linear confinement model and quadratic confinement model, resulting in different OGE coupling constants. The difference of the parameter $m_\sigma$ is then easy to be understood as it is partially needed to compensate the difference of the OGE interaction.

All though the parameters $g_u$, $g_s$ and $m_\sigma$ are different in linear confinement model and quadratic confinement model, one sees from Figs.~\ref{Fig:S-wave}-\ref{Fig:IJ-wave} that the $NN$ phase shifts from these two models are very close. In Fig.~\ref{Fig:OGE-H-S-wave} we show the diagonal matrix elements of OGE and the total Hamiltonian in the generator coordinate method (GCM) calculation for $NN$ $^1S_0$ and $^3S_1$ partial waves, which can be qualitatively regarded as the effective OGE and Hamiltonian between two clusters. In this figure, the red solid and blue dashed lines represent the matrix elements for Hamiltonian in models with linear and quadratic confinement, respectively. The red dotted and blue dash-dotted lines correspond to the results for OGE in models with linear and quadratic confinement, respectively. The horizontal axis denotes the generator coordinate which can qualitatively describe the distance between two clusters. One sees that the interaction of OGE at short range is more repulsive in linear confinement model than that in quadratic confinement model, so is the total Hamiltonian. Nevertheless, these two models result in quite similar $NN$ phase shifts,
%Actually, we only see small differences in the $^1S_0$ partial wave, where the phase shifts from the quadratic confinement model are a lit bit more attractive than those form the linear confinement model. While for all other partial waves, the differences of the results from these two models are tiny and can be ignored.
implying that the calculated $NN$ phase shifts in the energy region considered are not that sensitive to the short range interaction. For deeply bound $BB$ states, one may see difference of the results from these two models. The OGE and Hamiltonian matrix elements in the present work are also different from those in earlier quark model calculations. This to some extent also indicates that the $NN$ scattering data alone cannot uniquely determine the parameters in a constituent quark model. To get more reliable quark-quark interactions for the extension to study other $BB$ systems in a parameter-free way, one should constrain the model by fitting as many data as possible. In this regard, the present work, which describes the energies of single baryon ground states, the binding energy of the deuteron, and the $NN$ scattering phase shifts up to a total angular momentum $J=6$ in a rather consistent manner, may be considered as a step in the right direction.

Using the present model to get predictions for other $BB$ systems is beyond the scope of the present work. But based on the parameters listed in Table~\ref{Table:para} and the baryon size parameters shown in Table~\ref{Table:B-mass} and Fig.~\ref{Fig:baryon_mass_size}, it is still possible to make some interesting arguments without any detailed calculations. As we all know, the sigma meson exchange contributes attraction to all baryon-baryon systems. Its mass, $m_\sigma$, determined in the present work is 608 MeV in linear confinement model and 625 MeV in quadratic confinement model, both higher than that used in Ref.~\cite{Dai:2003}, 595 MeV. For $NN$ system, the attraction from the $\sigma$ meson exchange is not reduced in the present work since nucleon has a smaller size parameter, $b_u\approx 0.47$ fm, compared with $b_u=0.5$ fm used in Ref.~\cite{Dai:2003}.  However, for decuplet baryons, the size parameters determined by the variational method in the present work are much bigger than $b_u = 0.5$ fm. It is then expected that a much smaller attraction from the $\sigma$ meson exchange will be obtained in the present model when study the interactions of decuplet baryons, e.g. $\Delta\Delta$ or $\Omega\Omega$. Detailed investigations of these systems in the present model are in progress, and the results will be reported subsequently. 

In the present work, the short-range quark-quark interaction is dominated by OGE and quark exchange effects. In the literature, there are authors who studied the short-range interaction as stemming from vector-meson exchanges on quark level \cite{Dai:2003,Shimizu:2000}. As mentioned in Ref.~\cite{Huang:2005}, it is still a controversial and challenging problem whether OGE or vector-meson exchange is the right mechanism or both of them are important for describing the short-range quark-quark interaction. We postpone the study of this issue by incorporating the vector-meson exchanges in the quark-quark interaction Hamiltonian to our next work.

\section{Summary and conclusions} \label{Sec:summary}

The aim of the present work is to perform a consistent description of the masses of single baryons and the $NN$ scattering data by using the same Hamiltonian in a chiral SU(3) quark model. The considered quark-quark interactions includes, besides  the OGE potential and the phenomenological confinement potential, the potentials from the nonet scalar meson exchanges and the nonet pseudo-scalar meson exchanges derived from the spontaneous SU(3) chiral symmetry breaking. The masses of single baryons are calculated by using the Gaussian trial wave functions with the size parameters determined by the variational method, which ensures that all single baryons are minimums of the model Hamiltonian. The $NN$ interaction is dynamically investigated by using the resonating group method. It is found that the calculated masses of the octet and decuplet baryon ground states, the binding energy of the deuteron, and the $NN$ scattering phase shifts up to a total angular momentum $J=6$ and mixing parameters for the relevant coupled partial waves are in satisfactory agreement with the experiments. The present model may serve as a good starting point to achieve a consistent and unified description of the single baryon properties and the baryon-baryon dynamics. Investigations of baryon spectroscopy and the $BB$ interaction dynamics for systems other than $NN$ in a completely parameter-free way within the present model are planned for our future work.

\begin{acknowledgments}
The authors thank the discussions with Profs. Z. Y. Zhang, P. N. Shen, Y. B. Dong, P. Wang and J. X. Wang. This work is partially supported by the National Natural Science Foundation of China under Grants No.~11475181, No.~11635009 and No.~11475015,  the Youth Innovation Promotion Association of Chinese Academy of Sciences under Grant No.~2015358, and the Key Research Program of Frontier Sciences of Chinese Academy of Sciences under Grant No. Y7292610K1.
\end{acknowledgments}

\end{document}